\documentclass[prd,aps,twocolumn,showpacs]{revtex4}

\usepackage{amsfonts}
\usepackage{amsmath}
\usepackage{graphicx}
\begin{document}
\title{A New Approach to Time Domain Classification of Broadband Noise in Gravitational Wave Data}
\author{S. Mukherjee}
\author{P. Rizwan}
\author{R. Biswas}
\affiliation{Center for Gravitational Wave Astronomy, Dept. of Physics and Astronomy, The University of
Texas at Brownsville, 80 Fort Brown, Brownsville, TX 78520, U.S.A.}
\email[Corresponding author:]{soma@phys.utb.edu}

\begin{abstract}
Broadband noise in gravitational wave (GW) detectors, also known as triggers, can often be a deterrant to the efficiency with which astrophysical search pipelines detect sources. It is important to understand their instrumental or environmental origin so that they could be eliminated or accounted for in the data. Since the number of triggers is large, data mining approaches such as clustering and classification are useful tools for this task. Classification of triggers based on a handful of discrete properties has been done in the past. A rich information content is available in the waveform or `shape' of the triggers that has had a rather restricted exploration so far. This paper presents a new way to classify triggers deriving information from both trigger waveforms as well as their discrete physical properties using a sequential combination of the Longest Common Sub-Sequence (LCSS) and LCSS coupled with Fast Time Series Evaluation (FTSE) for waveform classification and the multidimensional hierarchical classification (MHC) analysis for the grouping based on physical properties. A generalized k-means algorithm is used with the LCSS (and LCSS+FTSE) for clustering the triggers using a validity measure to determine the correct number of clusters in absence of any prior knowledge. The results have been demonstrated by simulations and by application to a segment of real LIGO data from the sixth science run. 

\end{abstract}
\maketitle


\section{Introduction}
\label{introduction}

Gravitational-Wave (GW) detectors viz., LIGO~\cite{LIGO}, Virgo~\cite{Virgo} have been online in data-recording mode since 2000. LIGO has concluded its sixth science run (S6) in 2010.  Data have been archived from not only the GW channels, but also from several hundreds of auxiliary and environmental channels. There are two main broad categories in
which data analysis has been organized: astrophysical searches~\cite{Burst, I, P, SB} and detector characterization~\cite{DetChar}. These two tasks are not entirely independent. Detector characterization research products prepare the ground work for understanding the underlying noise and feed the astrophysical searches with information~\cite{DQFlag} that symbolize which data segments are relevant for GW search.

Data from the GW detectors have both broadband and narrowband noise. The narrowband noise (aka lines) is extensively studied and several methods~\cite{kalman, schutz+sintes} have been implemented to efficiently remove them from the data. However, reduction of the broadband noise (aka triggers) is a difficult problem that has not been explored to the fullest yet. The sensitivity of the detectors has improved steadily over the years~\cite{sensitivity}. With each step towards a more sensitive instrument, many new sources of noise have also been unearthed. Occurence of triggers in the data is a function of the operating conditions. Thus it is important to track down the sources to make the output data as high quality as possible and to reduce probability of false alarms. A very large effort has been undertaken to analyze the noise transients in the GW channels (referred to as DARM\_ERR channel) in the LIGO detectors \cite{multi-class06}. These consist of looking at triggers in time frequency planes \cite{qscan}, exploring loudest triggers seen in burst pipelines \cite{blocknormal, kleinewelle}, studying low frequency seismic disturbances \cite{mnft, rstone+mukherjee}, looking at specific types of triggers e.g. from photodiodes \cite{DetChar} and exploring structures present in the trigger population in dimensions higher than the usual three dimensional cartesian system using multidimensional hierarchical classification (MHC) methods \cite{multi-class06, multi-class07, mukherjee_lcss}. These methods are complementary and cast light on different aspects of the triggers seen in the data. 

GW data are archived in the format of discretely sampled time series from the main GW channel as well as from several hundreds of instrumental and environmental channels that are recorded specifically to monitor functioning of different instrumental subsystems and environmental activities that affect the GW channel data. The triggers arrive at a high rate in all channels. This requires data mining methods to keep up with near realtime information and to process the enormous volume of data for information extraction. Classification is the most effective way of addressing this problem. The methods of classifying large data sets in multidimensional parameter space bring an immediate reduction in the dimensionality of the problem under the assumption that existing classes show some common collective properties. In the context of triggers seen in the GW data, we would like to explore how many different classes of triggers are present and how to characterize these classes in terms of their origin. Development of a knowledge base in understanding the properties of the triggers thus seen in GW data contributes towards development of realistic noise models that are essential in proper assessment of peformance of astrophysical search pipelines.

There are several analysis pipelines that operate online on LIGO data to look for burst-like signals or triggers e.g. kleine welle (KW)~\cite{kleinewelle}, omega (OP)~\cite{omega} and waveburst
(WB)~\cite{WB}. The KW pipeline works on multiple channels -  the GW channel and several hundreds of auxiliary and
environmental channels. The threshold is kept such that the pipeline picks up triggers of all types at a steady rate. Unsupervised data mining methods like the
MHC analysis~\cite{multi-class06, multi-class07} has been developed and LIGO science data
have been analyzed in recent past. The aim of these studies has been to classify the population of triggers
seen in the GW, environmental and auxiliary channels into statistically significant distinct groups with uniform characteristics. These studies have been mostly based on a handful of discrete properties of the triggers viz. duration, central frequency and signal-to-noise ratio (snr). However, an important aspect of the triggers, viz. the `shape' factor has largely been overlooked. Shape of the triggers, or the waveform, often contains rich informaton. Temporal classification ~\cite{DTW, Agarwal, Pfoser, Saltenis} methods e.g. S-means and Constraint Validation~\cite{CVClust} have been developed in the recent past and studied on simulated GW data. However, they have not been tested on data from real GW detectors.

Another successful and often applied method is based on distances calculated using the Longest Common Subsequence (LCSS)~\cite{LCSS_vlachos}. The technique has been studied in GW data for the first time in a recent publication~\cite{mukherjee_lcss}. Preliminary results demonstrated have shown production of trigger clusters with similar shapes. 
The current paper explores fast, accurate and efficient methods for unsupervised classification of trigger
waveforms further. An analysis pipeline has been constructed based on LCSS and also LCSS in conjunction with Fast Time Series Evaluation (FTSE)~\cite{MorsePatel}. The latter is done to explore possibilities of increasing computational speed when very large datasets of triggers are involved. This is the first exploration of combined LCSS and FTSE methods in the context of GW data analysis. A second stage involving the MHC methods~\cite{multi-class06, multi-class07} is carried out to check the homogeneity of the clusters in the parameter space of their physical properties. This results in further segmentation of the triggers to appropriate them to their sources - instrumental or environmental. Some of the specific questions we explore in the paper are (i) Are LCSS and LCSS+FTSE suitable methods for fast and accurate trigger classification? (ii) How are the resulting classes of triggers characterized? (iii) What are the computational costs involved? (iv) How robust are these methods for GW trigger classification? (v) How can the analysis give relevant information for tracking down sources of non-GW triggers?  

It is shown in this study that application of the LCSS (or LCSS+FTSE) followed by MHC is a very useful and productive way to classify triggers based on their waveforms and characteristic physical properties. As has been found in the study with S6 data, the end result of the pipeline produces trigger classes with similar waveforms and amplitude, central frequency, Q-factor and snr range. Each of these classes of triggers is shown to be related to a group of auxiliary and environmental channels indicating the most probable sources of their origin. This combined classification pipeline performs better than methods based either only on discrete trigger properties or on trigger waveforms alone. Thus it can be turned into an effective, low latency trigger identification tool for GW detector characterization.

This paper is an illustration of the method and its advantages. Results from the sample S6 data chosen over a two day period are used to show how the method could be applied in science runs to extract information complementary to and in conjunction with the existing methods with low latency. The final outcome of this analysis shows existence of several statistically significant classes of triggers with distinct waveforms and physical properties coming from the GW channel in the test data set from S6. Post-classification analysis explores the couplings of these trigger classes to different sets of auxiliary and environmental systems. Application of LCSS and LCSS+FTSE (which yields classes based on waveforms) alone would give 19 distinct classes, while application of only the MHC analysis would have given 3 statistically significant classes. Thus, the proposed analysis clearly has advantage in using a bigger parameter space leading to a finer classification structure that helps in the identification of triggers by maximum utilization of its information content. Since each of the subgroups contains triggers with very characteristic properties related to a specific set of channels, the method proves useful in the classification of triggers seen in GW data and in helping with tracking down the sources or origins of the triggers. As a direct application to detector characterization, we can classify the triggers seen in GW science data into different groups with characteristic properties, related to specific instrumental and environmental sources. We can thus study the trend of various kinds of triggers and gain insight into how some of the channels may be reponsible in production of specific types of triggers.

The paper is organized as follows. Sec.~\ref{KW} describes the trigger generation process and
Sec.~\ref{algo} describes the FTSE and LCSS algorithms. We explain the pipeline for generation of trigger clusters in Sec.~\ref{pipe}.  Sec.~\ref{results} then presents results from numerical simulations and applications to S6. Conclusions and directions to future work are presented in Sec.~\ref{future}.


\section{Techniques for detecting triggers in LIGO data}
\label{KW}

Techniques for detection of burst-like triggers in the GW instrumental output data stream are described concisely in an earlier paper ~\cite{kleinewelle}. In general, such techniques project the data onto a basis that spans the parameter space of the burst-like signals. A measurement becomes optimal  when there is an exact match between a member in the contructed basis and a burst. The snr $\varrho$ in this case is defined as
\begin{equation}
\varrho^2 = \frac{||h||^2}{S_h (f)},
\end{equation}
where $||h||^2$ is the total energy content of the signal and $S_h (f)$ is the one-sided power spectral density. In cases where a close match between the signal and the member of the basis is not achieved, the bursts are characterized by a quality factor $Q$ defined as 
\begin{equation}
Q = \frac{f_c}{\sigma_f}
\end{equation}
where $f_c$ is the central frequency and $\sigma_f ^2$ is the bandwidth. The time-frequency plane is thus titled by the $Q$ values where the signals are represented by localized pixels with the same $Q$.



\section{Algorithms for trigger Classification in a Multidimensional Space}
\label{algo}

There are two main issues that one must address while developing an efficient feature-based classification method – (i) the data may have temporal gaps, i.e. there may be same pattern occurring at different time epochs and (ii) the classification preferably should be automatic (to keep up with the online feedback systems) and thus unsupervised classification methods that are {\it robust to noise} need to be explored. 

\subsection{Longest Common Sub Sequence}
Let us first take a look at why LCSS algorithm is efficient and how it fits into GW data analysis. In any unsupervised classification method, the first step is to calculate the distance between points in the parameter space. Even though Euclidean distance is the most commonly used method, it is not suitable for addressing the first difficulty mentioned above. Thus, two triggers that have the same waveform may show a high Euclidean distance if they are not occurring simultaneously. This will be considered a redundant cluster from the physical point of view. LCSS algorithm is able to compute a match between two time series by calculating metrics for triggers that do not necessarily occur at the same time without having to rearrange the sample sequences to coincide. 

As an example, let us consider a {\it sequence} of characters $x_m, x_n, x_p, x_p, x_q, x_n, x_q$. A {\it subsequence} is defined as a set of characters that appear in an order from the left to the 
right, but not necessarily consecutively. Thus, [$x_m, x_n, x_p$], [$x_m, x_p, x_p, x_q$], [$x_p$], [$x_m, x_n, x_p, x_p, x_q, x_n$] are subsequences, but [$x_p, x_p, x_n$] is not a subsequence. A {\it common subequence} of two sequences is a subsequence that appears in both sequences. A {\it longest common subequence} (LCSS) is a common subsequence of maximal length. For example, suppose
\begin{equation}
s_1 =   uu {\rm \bf uvv}w{\rm \bf t}w{\rm \bf uw}t{\rm \bf tu}tt{\rm \bf v} w{\rm \bf tt} v {\rm \bf t} u {\rm \bf w} uu \\
\end{equation}
and 
\begin{equation}
s_2 = v{\rm \bf u} vv{\rm \bf vvtu} uw{\rm \bf wtuv} v{\rm \bf tttw} wttv,
\end{equation}
an LCSS (denoted by LCSS$(s_1, s_2)$) is given by $ \rm \bf {uvvtuwtuvtttw}$. The algorithm operates on the principle of enumeration all subsequences of $s_1$, followed by checking if they are subsequences of
$s_2$ as well.

Let us now look at the theory of LCSS in the context of the trigger waveforms in the present study. 
Formally, this amounts to comparing two input trigger time series sequences $ X(1 \dots m)$ and $Y(1 \dots n)$, where $m$ and $n$ denote the length of the sequences $X$ and $Y$ respectively.  
The length of LCSS of $X$ and $Y$ (or written as LCSS$(X, Y)$) will be denoted by $ \zeta $. The recurrence relation~\cite{bergroth} leading to the length of the LCSS for each pair $[X(1 \dots i), Y(1 \dots j)]$ is given as follows:

\begin{equation}
\alpha(i, j)= 0                                
\end{equation}
if $X$ or $Y$ is empty sequence, i.e. if $i = 0$ or $j = 0$; 
\begin{equation}
\alpha(i, j) = 1+ \alpha(i - 1, j - 1)
\end{equation}
if $X(i) = Y(j)$;             
\begin{equation}
\alpha(i,j) = max \{ \alpha(i -1, j), \alpha(i, j - 1) \}
\end{equation}
if  $X(i) \ne Y(j)$, 
where 
\begin{equation}
\alpha(i,j) = \zeta([X(1 \dots i), Y(1 \dots j)]).
\end{equation}

\begin{figure}
\includegraphics[scale=0.45]{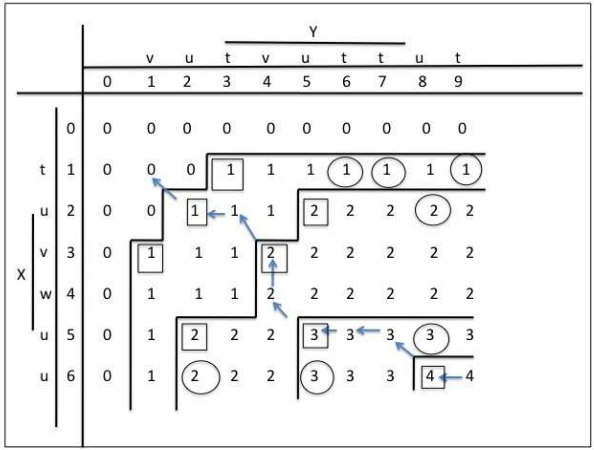}
\caption{
\label{lcssfig}
The figure shows graphically how the LCSS algorithm works. The two-dimensional array with the two sequences $X$ and $Y$ is shown along the two axes. Initially the cells in the array have uniform entries of zeros. In the next step, we look for an element to element match. Whenever a match is found, the cell entry in incremented by one.  The sequence of arrows show a possible LCSS path. The LCSS path in this case indicates that there is a match of four elements in the sequence viz., ${\bf uvuu}$~\cite{bergroth, cormen}. More details are given in \ref{algo}.
}
\end{figure}

The algorithm is shown graphically in figure \ref{lcssfig}. A single $\alpha$ value is localized in the sense that it depends only on the three neighboring values. After the table has been filled, the length of the subsequence is found in 
\begin{equation}
\alpha(m,n) = \zeta([X(1 \dots m), Y(1 \dots n)]).
\end{equation}
The common subsequence is found by backtracking from $\alpha(m,n)$ by following at each step, the pointers that are set during the calculation of the values. When a match is found, the LCSS is upgraded. In this way, one can traverse a path through the LCSS table (as shown in \ref{lcssfig}) until a length of zero is reached.  In this case, $\zeta(X,Y)$ is =4 and the LCSS corresponding to the path shown is {\bf uvuu}. 

The algorithm works as follows. A pair $(i,j)$ defines a match if 
\begin{equation}
X(i) = Y(j).
\end{equation}
The set of all matches is given by 
\begin{equation}
\eta = \{ (i,j) | X(i) = Y(j), 1 \le i \le m, 1 \le j \le n \}. 
\end{equation}
Each match belongs to a class
\begin{equation}
\Omega_k= \{ ([i,j] | [i,j] \in \eta ; \alpha(i,j) =k \}, 1 \le k \le \zeta.
\end{equation}
A match belonging to the $\Omega_k$ is called a $k$-match. In the figure \ref{lcssfig}, the marked entries  define the class $\Omega_k$. Since each match belongs to only one class, these classes partition all matched of $\eta$.

Some $k$-matches are more useful algorithmically (e.g. square marks in the figure \ref{lcssfig}) than the others (e.g. circle marks in the figure \ref{lcssfig}) . This can be proven as follows. Let us consider matches [$i,j$] and  [$i ^\prime ,j ^\prime $] of $\Omega_k$ for $i = i ^\prime$ and $j \le j ^\prime$ or $ i \le i ^\prime$ and $j = j ^\prime$. Every element of $\Omega_{k+1}$ that follow [$i ^\prime ,j ^\prime $] should also follow [$i,j$] in the LCSS. Thus, it is sufficient to consider only the dominant matches [$i,j$]. Let $\varphi_k$ be the set of all dominant matches. The regions in the figure where $\alpha(i,j)$ values are equal (shown by vertical and horizontal lines) are called LCSS contours. Each $k$-match lies immediately below the $k^{th}$ contour. These contours are defined by an ordering property. If  $\varphi_k$ = $[i_1, j_1], [i_2, j_2], \dots , [i_l, j_l, ]$, the matches can be renumbered such that $i_1 < i_2< \dots  <i_l$ and $j_1 > j_2 > \dots > j_l$. The strategy for locating the dominant matches is based on advancing from contour to contour.

\subsection{K-means}

Once the distances are calculated according to equation (5), a generalized k-means [29] algorithm is employed to form the clusters. The reason as to the choice of k-means is dictated by the fact that this allows formation of homogeneous clusters that are insensitive to outliers. 

K-means~\cite{k1,k2} uses two parameters to start with – the number of clusters $K$ and the set of elements $ D = [t_{1}, t_{2},\ldots,t_{n}]$. The algorithm works as follows.

Let $ M = [m_{1}, m_{2},\ldots,m_{k}]$ be the set of centroids assigned randomly and the size of the set $M$ equals $K$. Each item $t_{i}$ is placed into the cluster which has the nearest mean. $M$ is populated with a new value of mean for each cluster $K_{i}$. The cluster mean of $K_{i} = [t_{i1}, t_{i2},\ldots,t_{il}]$ is usually (but not necessarily) calculated as:\[m_i=\frac{1}{l}\sum_{j=1}^l t_{ij}\]\newline
where $l$ is the number of items and $t_{i}$ is the item placed in each cluster $K_{i}$.\newline
The last three steps are repeated until a defined convergence condition is met such as no further change in the membership of the clusters.\\







One of the short falls of k-means is that the algorithm needs a predefined value of K or the number of clusters, which is exactly what the proposed clustering technique aims to know. In case of unknown data under test, the number of significant groups in the classification structure is unknown and hence it cannot be supplied to K-means.

Several validity measures have been developed to determine the value of K in k-means.  Using the ratio of intra cluster distances $\delta_{intra}$ to inter cluster distances $\delta_{inter}$, a simple validity measure $V_{min}$ is used to find out the optimum number of clusters~\cite{SidTuri}. The time series representation of the triggers within a cluster ( a {\it point} in the multivariate parameter space) must be as similar as possible and similar points belonging to different clusters must be as different as possible (in the same multivariate parameter space) to ensure compactness of clusters. Therefore, the intra cluster distance should be minimum and the inter cluster distance should be maximum. Since,

\begin{equation}
V_{min} = \frac{\delta_{intra}}{\delta_{inter}}
\end{equation}

 the value of K which makes the validity measure minimum, is the ideal one. In practice it is expected that the values of K found from the above method would vary slightly on different runs. This happens because, in the k-means algorithm, the centroids are assigned randomly. Therefore, the validity measure was computed one hundred times on the same sets of data and the most frequent value of K is chosen.

The role of LCSS in this clustering scheme is to generate an adjacency matrix containing information about the pairwise distances between trigger time series. Once the matrix is created, it can now be supplied to K-means which treats it as an ordinary input and operate on it based on the algorithm.

\subsection{Attempt to Improve Computational Speed: Fast Time Series Evaluation}


If we need to implement the classification algorithm as a near real-time tool for trigger identification, we need to make the process computationally fast and efficient. To address this question, we also investigated methods that could enhance the computational  speed of the LCSS algorithm. We thus investigate the Fast Time Series Evaluation algorithm (FTSE)~\cite{MorsePatel} in an attempt to make LCSS faster. LCSS in conjunction with FTSE claims to be faster than LCSS algorithm alone because unlike LCSS calculations, FTSE does not use a two dimensional array to compare matches between two time series. Values in one time series are entered into a grid. Then a point in the other time series probes into its respective grid cell (of the same grid) to check if points of the first time series reside there. The construction of the grid ensures that if points are found residing in that grid cell, they must match the probing point within a defined threshold. To put it simply, in LCSS computed with FTSE, the comparison between two time series occurs {\it only} between the ‘intersecting’ portions  and that is the underlying reason why LCSS computed with FTSE is expected to be faster than LCSS - where the comparison occurs throughout the entire length of both time series. 

\begin{figure}
\includegraphics[scale=0.5]{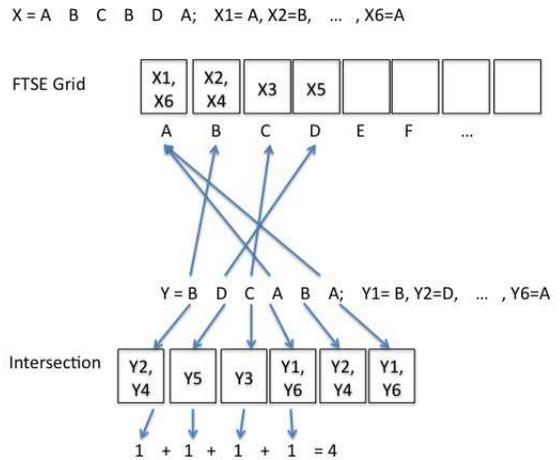}
\caption{
\label{Dyn_FTSELCSS}
The figure shows graphically how the LCSS +FTSE algorithm works. The top row represents the cells of the grid. According to the figure, indices of the matching elements of a series $X$ are put into corresponding cells of the grid. The elements of the second series $Y$ are compared with the grid cells formed from $X$ to check if points of the first time series reside there. The construction of the grid ensures that if points are found residing in that grid cell, they must match the probing point within a defined threshold. In this case, the element $B$ of $Y$ is found to match the second grid cell and thus the value inside the cell enters the intersection list. Likewise, the other elements of $Y$ are also probed in a similar manner and the total matching points between the sequences are recorded in the intersection list~\cite{MorsePatel}. In this example, the contributing indices 2, 3, 4, 6 gives the subsequence, viz., $B, C, B, A$. 
}
\end{figure}

The average cost of calculating LCSS is $O(p\times q)$ where $p$ and $q$ are the lengths of two sequences being compared. The average cost of FTSE computing LCSS is $O(M^\prime+Lq)$ where $M^\prime$ is the number of matches, $L$ is the longest intersection between the two sequences and $q$ is the length of the probing sequence [3]. The process of computing LCSS with FTSE is elaborately shown in figure \ref{Dyn_FTSELCSS} . The top row represents the cells of the grid. According to the figure, matching elements of a series $X$ are put into the same cells of the grid. The elements of the second series $Y$ compared with the grid cells formed from $X$ to check if points of the first time series reside there. The construction of the grid ensures that if points are found residing in that grid cell, they must match the probing point within a defined threshold. In this case, the element $B$ of $Y$ is found to match the second grid cell and thus the length of the matching subsequence is augmented by one. Likewise, the other elements are also probed in a similar manner and the total matching length between the sequences are recorded. Once all the matching lengths of the trigger signals are calculated, the signals are clustered using the method described above.

\subsection{Multidimensional Hierarchical Classification algorithm}
\label{multipipe}
The MHC pipeline implements a hierarchical algorithm [14-19] that applies a variance minimization criterion and groups together burst triggers detected by the KW [11] pipeline based on their similarity in the higher dimensional space spanned by properties like the trigger duration, frequency, snr and statistical significance. 

The algorithm starts with calculation of a Euclidean distance between vectors $X_m$ and $X_n$ defined by, 
\begin{equation}
d^2_{mn} = \Sigma_i (X_m - X_n)^2_i      
\end{equation}                              

The data matrix has dimension $p \times q$ , where each of $p$ triggers is described
by $(1 \times q)$ vectors $X_1, X_2, \dots , X_p$. Calculation of distance is followed by computation of suitable measure of proximity between two groups of objects. These are called `linkage' criteria. We adopt the criterion of `complete' linkage which measures the largest distance between objects in the two clusters. If $N_m$ is the number of objects in class $m$ and $N_n$ is the number of objects in class $n$, and $X_{mj}$ is the $j$th object in class $m$, a complete linkage is defined as follows.
\begin{equation}
 d(m, n) = max(\Delta(X_{mj}, X_{nk})),     
\end{equation}

with $j$ ranging between $ 1, \dots,N_m$ and k ranging between $1, \dots ,N_n$.  $\Delta$ denotes the distance. 
This stage of the algorithm results in a hierarchy from $p$ clusters with one object to one cluster with $p$ objects. The choice of significant clusters is given by computation of the correlation coefficient, $`r'$~\cite{pearson1896}. $r^2$ is related to the fraction of the total variance accounted for partitioning into $s$ clusters. $r^2$ is defined as follows. 
If 
\begin{equation}
W_i = \Sigma_{j=1} ^{N_i} |X_j - \bar{X} |^2 ,
\end{equation}
\begin{equation}
r^2 = 1 - \Sigma_{i=1} ^s \frac{W_i}{\Sigma_{l=1} ^p |X_l - \overline{X_{total}} |^2}
\end{equation}
$N_i$ denotes the number of members in the $i^{th}$ class, $ \bar{X}$ denotes the unweighted mean of the population in the $i^{th}$ class and $\overline{X_{total}} $ denotes the mean of the entire population~\cite{manova, gamma}.

The statistical significance of the classification scheme can be verified by using the method of Multivariate Analysis of Variance (MANOVA)~\cite{manova}. Assuming that the underlying distribution is a multinormal mixture~\cite{gamma, johnson}, the model is given below. For an $n$ dimensional data set with $m$ clusters each with $p_k$ members, the $i^{th}$ trigger in the $j^{th}$ cluster gives an $n$ dimensional vector
\begin{equation}
X_{ij} = \mu + \tau_j + \epsilon_{ij}.
\end{equation}
$\mu$ is the population mean of the total population, $\tau_j$ is the offset of the $ j^{th}$ cluster mean from $\mu$ and $\epsilon_{ij}$ is the scatter of the points around the eman value. The hypothesis to be tested is as follows.
\begin{equation}
H0: \tau_1 = \tau_2 = \tau_3 = \dots \tau_m = 0.
\end{equation}
Let the sum of squares be written as
\begin{equation}
SS=\Sigma_{k=1} ^ m \Sigma_{j=1} ^{p_k} (\overline{X_{k}} - \overline{X})(\overline{X_{k}} - \overline{X})^T
\end{equation}
and the cross products are written as
\begin{equation}
CP=\Sigma_{k=1} ^ m  \Sigma_{j=1} ^{p_k} (\overline{X_{kj}} - \overline{X_k})(\overline{X_{kj}} - \overline{X_k})^T,
\end{equation}
The test statistics are as follows:\\
(i) Wilks lambda~\cite{wilks}:
\begin{equation}
\lambda^* = \frac{det(CP)}{det(SS + CP)}
\end{equation}
(ii)Pillais trace~\cite{pillai}:
\begin{equation}
V = trace[SS \times (SS + CP) - 1]
\end{equation}
(iii) Hotelling Lawley's trace (or Mahalanobis $D^2$ statistic)~\cite{mahalanobis}
\begin{equation}
U = trace(CP^{-1} \times SS).
\end{equation}
All the test statistics follow the non-central $F$ distribution ~\cite{fstat}.


\section{Analysis Pipeline}
\label{pipe}
The analysis is carried out in three main stages - (a) using simulated triggers without additive noise; (b) using simulated triggers with additive Gaussian white noise and (c) using a segment of LIGO S6 data chosen over a two day period.

\subsection{Simulated triggers without additive noise}

\begin{figure}
\includegraphics[scale=0.6]{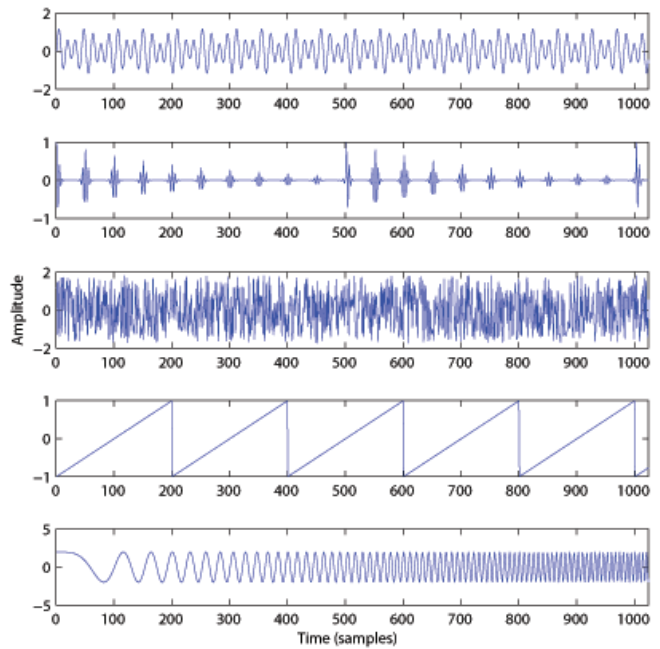}
\caption{
\label{sim_no_noise}
Five different types of simulated trigger waveforms were used in the classification pipeline with LCSS and LCSS+FTSE. These waveforms are (from top to bottom) mixture sinusoids, pulse trains, noise generated with a low order ARMA model (\cite{ARMA}), a simple triangular sawtooth wave and a chirp. The simulated set consisted of 760 waveforms with these shapes but varying in amplitude, frequency, relative location on the time axis etc. The figure represents typical examples of each type. No noise was added to these trigger waveforms. The x-axis denotes time samples and the y-axis denotes amplitude. 
}
\end{figure}

We first generate a data set of 760 simulated waveforms with variable parameters, each 1024 samples long. The waveforms are in the shapes of Mixture sinusoids, Pulse trains, noise generated with a low order Auto Regressive Moving Average (ARMA \cite{ARMA}) model, Triangular sawtooth and chirps with varying parameters, i.e. varying amplitude, frequency, width, location on the time axis etc. These waveform models are selected to generate waveforms of diverse nature and shapes. The motivation behind including ARMA-model based waveforms in the simulation is because it is a general scheme that can model various different types of waveforms (up to second moment), including the type of outputs we see in our GW detectors. The amplitude is normalized. The waveform database is introduced to the pipeline as the prime input. The waveforms are shown schematically in figure \ref{sim_no_noise}. The number of clusters is determined by k-means.

\subsection{Simulated triggers with additive Gaussian noise}

\begin{figure}
\includegraphics[scale=0.6]{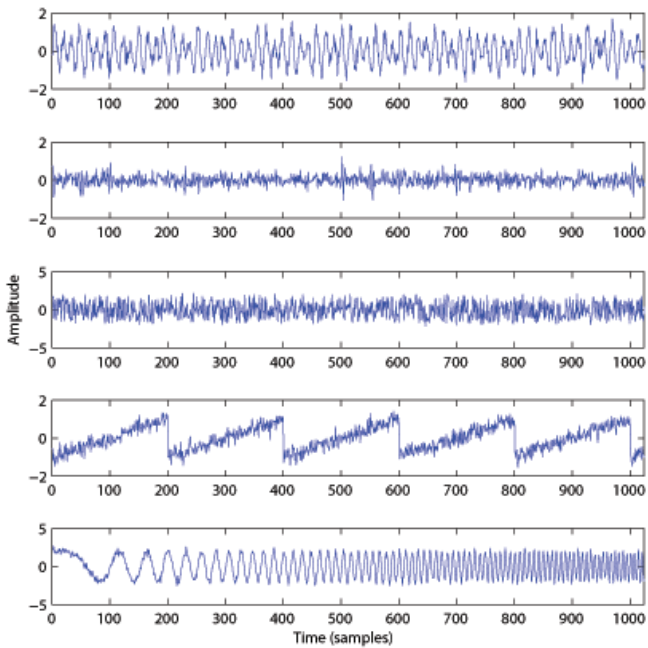}
\caption{
\label{sim_noise_trigger}
This figure represents simulated trigger waveforms with added Gaussian white noise used in the classification pipeline with LCSS and LCSS+FTSE. These waveforms are (from top to bottom) mixture sinusoids, pulse trains, noise generated with a low order ARMA model, a simple triangular sawtooth wave and a chirp. The figure represents typical examples of each type. The x-axis denotes time samples and the y-axis denotes amplitude. The standard deviation of the noise added to the trigger waveforms shown in the current figure is $\sigma=0.25$.
}
\end{figure}

In the next step of this study, we generate a data set of 760 simulated waveforms of five different types of shape with variable parameters, each 1024 samples long, as described in the previous section. The amplitude is normalized. Each waveform data stream is mixed with Gaussian white noise. The output thus is a noisy waveform, as shown in figure \ref{sim_noise_trigger}. The snr of the triggers are kept in a range of 2 and 20. This is fed to the classification analysis pipeline. As before, the number of clusters is determind by generalized k-means. 

\subsection{
\label{L4}
LIGO sixth science run trigger database} 

Having gained insight with the simulations, we now apply the analysis pipeline to classify triggers seen in LIGO S6 data. We have used triggers found in the Omega trigger catalog ~\cite{OmegaCatalog, omega, kleinewelle} during S6. Three hundred and forty Omega triggers from the GW channel seen in the test data set (with snr $>$ 12) are selected and subjected to the analysis pipeline. The aim of this exercise is to see if this scheme of classification can find statistically significant groups of triggers in this test population based on the shape of the waveforms of the triggers. Triggers in the Omega catalog are also described in terms of four discrete properties. These are central frequency, amplitude, snr and the quality factor (or Q-factor).

Once the classification of triggers take place, we address the important detector characterization question: (i) How to characterize these triggers and (ii) What are the possible sources of these triggers, i.e. how do they relate to the auxiliary and environmental channel triggers?
Class characterization and trigger identification steps are as follows.\\
(i) We take each trigger from a given sub-class that results from the main classification structure of the GW triggers and take an Omega scan ~\cite{OmegaScan} around the peak time of the GW trigger. Omega scans produce time-frequency plots of all auxiliary and environmental channel data that coincides with the trigger peak time. \\
(ii) A histogram is constructed to see the distribution of the occurrence of triggers in the auxiliary and environmental channels for all the time windows corresponding to triggers in a particular sub-class. The highest frequencies are recorded.\\
(iii) The existing data quality flags in the LSC detector characterization literature~\cite{DQFlag} are also noted for comparison. This also shows if this method points to newer auxiliary and environmental sources other than the existing data quality flags. 

{\em Data Conditioning}: The time series segments corresponding to the triggers are first subjected to conditioning. The following sequence of operations is executed. \\
(i)	Selection of triggers with snr greater than 12 from KW database;\\
(ii)	Extraction of raw GW channel data, centered around the trigger (extracted time series noted by, say, $ {q^\prime_i}.)$\\
(iii)	Whitening $ {q^\prime_i}$ ~\cite{whiten} and dynamically removing~\cite{kalman, schutz+sintes} the narrowband noise present in ${q^\prime_i}$. The resulting time series is denoted by ${cq^\prime_i}$.\\
(iv)	Filtering ${cq^\prime_i}$ with a bandwidth of $\delta f$ around $f_c$, where $f_c$ is the central frequency of the trigger as noted in the KW catalog. The resulting time series is denoted by ${fc_{q^\prime_i}}$.\\
(v)	Re-sampling ${fc_{q^\prime_i}}$ to represent the appropriate bandwidth.

\begin{figure}
\includegraphics[scale=0.45]{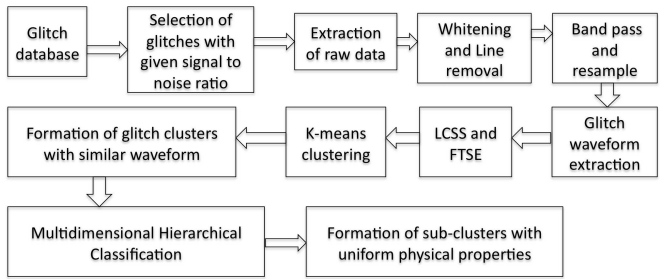}
\caption{
\label{pipefig}
The figure shows the full analysis pipeline as applied to a sample of S6 triggers in this study. 
 The trigger database is used to first select triggers with snr $>$ 12. The time stamps on the triggers are used to extract the corresponding time domain information or waveforms. The time domain data is conditioned elaborately to reduce noise that is mixed with the triggers i.e. first whitened and then narrowband noise is subtracted from the data. The conditioned time series are resampled and appropriately bandpassed to record the waveform. These are fed first into the LCSS pipeline and then independently into the LCSS and LCSS-FTSE combined pipeline.  The individual clusters thus generated are further subjected to the MHC pipeline for finer classification structures and post-processing. 
}
\end{figure}

\begin{figure}
\includegraphics[scale=0.45]{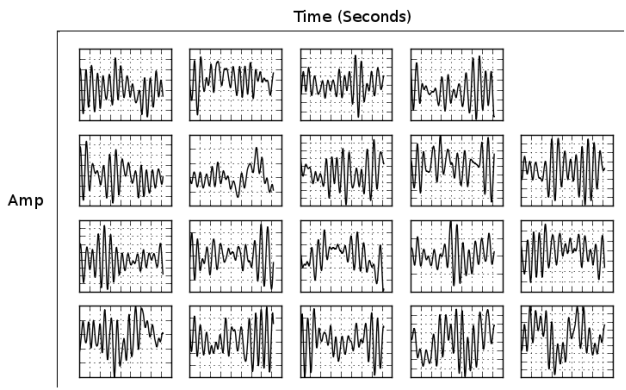}
\caption{
\label{Trigger_Matrix}
The figure represents samples of various types waveforms that were seen in the test data from LIGO's sixth science run. These trigger waveforms were obtained after application of the data conditioning part of the pipeline to the raw data. The x-axis represents time in arbitrary units and the y-axis represents amplitude also in arbitrary units. 
}
\end{figure}

As shown in the figure \ref{pipefig}, the trigger database is used to first select triggers with snr $>$ 12. The time stamps on the triggers are used to extract the corresponding time domain raw data. The raw data are conditioned elaborately to reduce noise that is mixed with the triggers i.e. they are whitened and then narrowband noise is subtracted from the data. The conditioned time series are resampled and appropriately bandpassed to record the waveform. Figure ~\ref{Trigger_Matrix} shows the different types of waveforms that have been found in the test data. These are fed first into the LCSS pipeline. The end product of the analysis is a set of individual uniform groups of trigers with similar shape parameters. The number of clusters is determind by generalized k-means. The individual clusters thus generated are further subjected to the MHC pipeline for finer classification structures that can be related to their most probable instrumental or environmental origins.

\section{Analysis Results}
\label{results}
 
\begin{figure}
\includegraphics[scale=0.6]{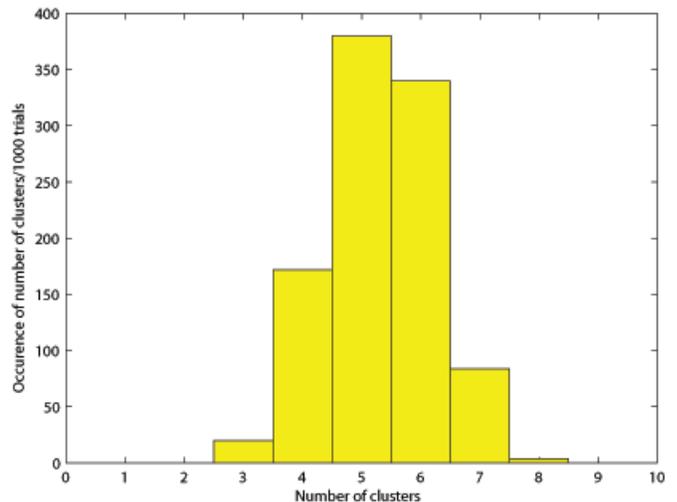}
\caption{
\label{sim_no_noise_hist}
The figure represents results of the classification structure. K-means is run on the simulated waveform database 1000 times and the number of classes deemed most significant (by the validity measure) is recorded. The histogram shows that in most of the cases, the data shows existence of five distinct classes, which is the true number of classes. The x-axis denotes the number of significant classes. The y-axis shows the corresponding number of trials. 
}
\end{figure}

\begin{figure}
\includegraphics[scale=0.4]{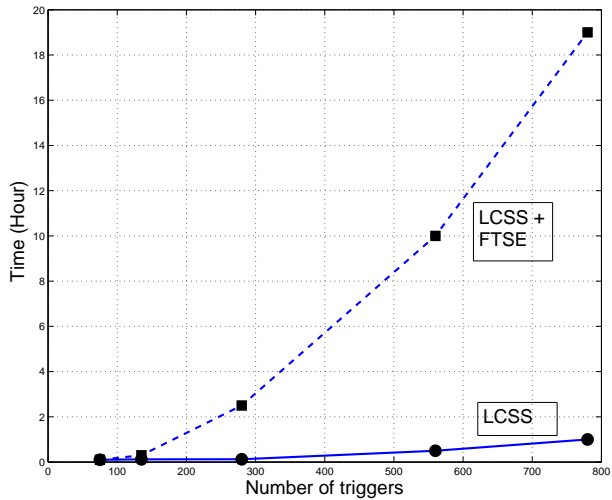}
\caption{
\label{sim_no_noise_compspeed}
The figure presents a comparison of computational speeds of the analysis pipeline using LCSS and combined FTSE with LCSS. While the two methods seem to have comparable computational speeds (i.e. not significantly different) up to $n < 120$, the LCSS+FTSE algorithm is found to be more computationally intensive beyond that point. The x-axis represents the number of triggers classified and the y-axis represents computation time in hours. 
}
\end{figure}

The results of the analysis are demonstrated in this section. Figure \ref{sim_no_noise}   shows the different simulated trigger waveforms that are subjected to the analysis without any additive noise. Figure \ref{sim_no_noise_hist}  shows the results of the classification structure. K-means (as described in section III) has been run on the database 1000 times and the number of classes deemed most significant (by the validity measure $V_{min}$) is recorded. The histogram shows a peak at 5 significant classes which was the true number of clusters in the data. Both LCSS and LCSS+FTSE showed identical classification structure. Figure \ref{sim_no_noise_compspeed} shows the computational speeds for LCSS and LCSS+FTSE. Contrary to hypothesis, LCSS+FTSE is found to be much more computationally intensive as the sample size grows larger than 120. The reason for this divergence is is explained in \ref{future}.

\begin{figure}
\includegraphics[scale=0.6]{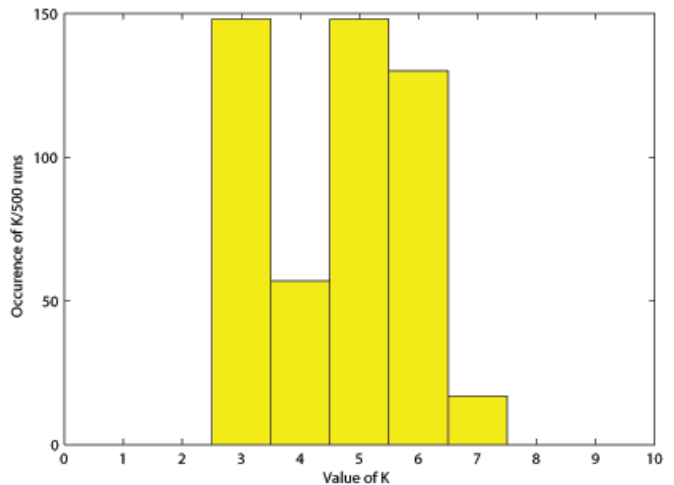}
\caption{
\label{sim_noise_hist2}
The figure represents results of the classification structure for the case of simulated triggers when the noise level is high (i.e. triggers with low snr). K-means was run on the database 500 times and the number of classes deemed most significant (by the validity measure) was recorded.  For $\sigma$=0.3, the histogram peaks at five clusters, but the overall shape of the histogram is broad, indicating that three or six clusters are equally probable. The results show an overwhelming majority of three clusters for cases with high noise 
($\sigma \ge  $ 0.3). More details are given in section \ref{results}. The x-axis denotes the number of significant classes. The y-axis shows the corresponding number of trials. 
}
\end{figure}

The next set of studies are conducted with the same set of waveforms, but now mixed with Gaussian white noise with varying standard deviation $\sigma$. The values of $\sigma$ varies from $0.1$ to $0.8$ in steps of $0.1$. Figure \ref{sim_noise_trigger} shows the typical trigger waveforms of various types with added noise. Classification is carried out by k-means as in the previous case. The best value of the number of clusters is determined by the validity measure $V_{min}$. It is found that the classification structure starts to deteriorate with increasing $\sigma$, i.e. decreasing signal to noise ratio (snr).  For $\sigma \ge 0.3$, the histogram peaks at five clusters, but the overall shape of the histogram is broad, indicating that three or six clusters are equally probable (\ref{sim_noise_hist2}). An overwhelming majority of three clusters for cases with high noise ($\sigma \ge  $ 0.3) is observed. The reason for the deterioration is easily explained. With increasing noise, many of the noise dominated waveforms, e.g. the mixture sinusoidal waveforms, the pulse train waveforms and the ARMA based waveforms look similar and are thus classified as one group.

\begin{figure}
\includegraphics[scale=0.45]{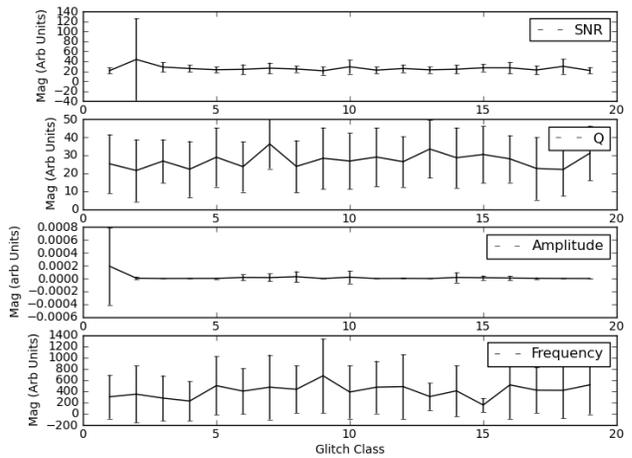}
\caption{
\label{ClusterProperties}
The figure shows variation of snr, amplitude, Q-value and frequency that describe triggers in the study. While the amplitude shows least variation, the other three properties have wide error bars indicating that the classes based on similar waveforms are heterogeneous. The x-axis represents time in arbitrary units. The y-axis represents the respective units of each property displayed.
}
\end{figure}

As stated earlier in this section, both LCSS and LCSS+FTSE showed identical classification structure, but LCSS+FTSE is found to be much more computationally intensive. Thus, we found that there is no real advantage at this stage to continue to apply LCSS+FTSE to further analysis. We thus continue the application to S6 sample data using only LCSS algorithm.

The analysis yielded 19 significant clusters of triggers. As mentioned above, figure ~\ref{Trigger_Matrix} shows typical waveform from each class. The shape of these waveforms forms the basis of classification into distinct clusters by the LCSS pipeline. Figure ~\ref{ClusterProperties} shows how the four discrete properties of the omega triggers (snr, amplitude, frequency and Q-value) vary between the different clusters. While the amplitude shows least variation, the other three properties have wide error bars indicating that the classes based on similar waveforms are heterogeneous. We further investigated if the clusters thus produced have any significant sub-clusters present in them. 

\begin{figure}
\includegraphics[scale=0.45]{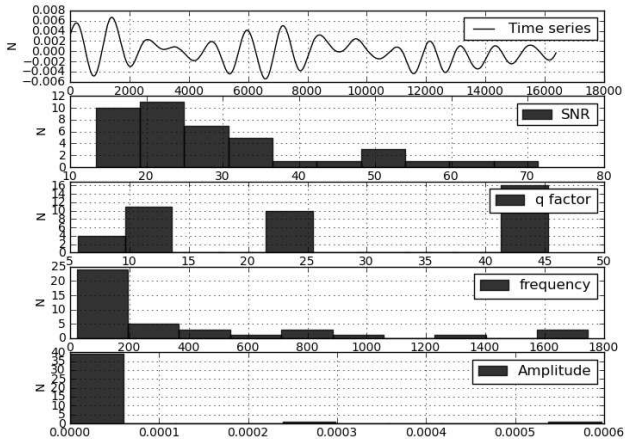}
\caption{
\label{Class10}
The top panel in this figure shows a typical example of trigger waveforms that is classified belonging to class \#10. The panels below the top panel show how some of the chief attributes viz., central frequency, the snr, the Q-value and the amplitude of the triggers belonging to this class are distributed. The x-axis represents each of the properties expressed in arbitrary units. The y-axis represents numbers.
}
\end{figure}

\begin{figure}
\includegraphics[scale=0.45]{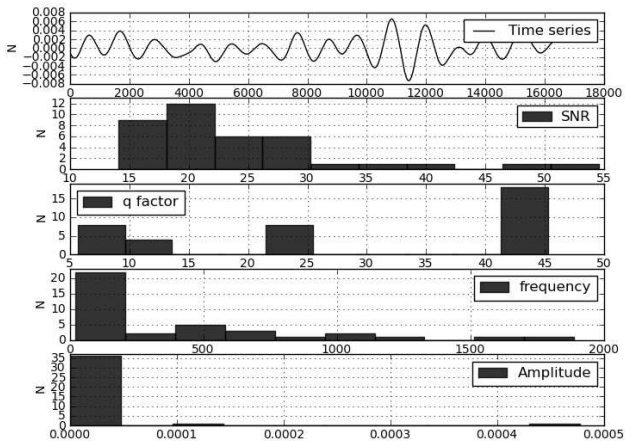}
\caption{
\label{Class14}
The top panel in this figure shows a typical example of trigger waveforms that is classified belonging to class \#14. The panels below the top panel show how some of the chief attributes viz., central frequency, the snr, the Q-value and the amplitude of the triggers belonging to this class are distributed. The x-axis represents each of the properties expressed in arbitrary units. The y-axis represents numbers.
}
\end{figure}

Figure~\ref{Class10} shows details of  the properties of one of the 19 classes of triggers found in the test data (class \# 10). The top panel in this figure shows a typical example of trigger waveforms that is classified belonging to this group. The panels below the top panel show how some of the chief attributes viz., central frequency, the snr, the Q-value and the amplitude of the triggers belonging to this class are distributed.  A very similar picture arises for another trigger class in the study~\ref{Class14} (class \#14). 

As it appears from the distributions of the properties like amplitude, central frequency, Q-value and snr for each group, the groups are fairly diverse within itself, even though they seem to indicate similar types of waveforms. This prompts a further look into the groups themselves. Here, we have implemented the MHC method, previously developed and described in detail in \cite{multi-class06, multi-class07}. The classification is based on the four dimensional space spanned by amplitude, central frequency, Q-value and snr of the triggers. Statistical significance of the classification structure thus found has been validated at $p < 10^{-6}$ level.

This sequential application of the LCSS and the MHC pipelines has proven very useful as far as distunguishing different types of triggers from various sources is concerned. The LCSS pipeline separates out the triggers with similarity of waveform, thus taking a first cut at fragmenting the huge parameter space into a finite number of partially uniform (in terms of waveform) groups. The MHC then further explores the finer physical property based groups present in these broader classes.

Thus, the final outcome of the analysis on the example data set in this study shows existence of 19 statistically significant classes of triggers with distinct waveforms and further classification based on physical properties, coming from GW channel and different sets of auxiliary and environmental systems leads to uncovering more groups or clusters of triggers. The combination of the two methods yield trigger clusters that would not appear by any one of the component algorithms - for example, application of LCSS (which yields classes based on waveforms) alone would give 19 distinct classes, while application of only the MHC analysis would have given at most 3 statistically significant classes (being constrained by the dimensional of the parameter space.) Since each of the subgroups contain triggers with very characteristic properties and can be related to a specific set of channels, the method proves useful in classification of triggers seen in GW data and in helping with tracking down the sources or origins of the triggers. 

As a direct application to detector characterization, we can classify the triggers seen in GW science data into different groups with characteristic properties, related to specific groups of channels. We can thus study the pattern trend of various kinds of triggers and gain insight into how some of the channels may be reponsible in production of specific types of triggers.

\subsection{Post classification analysis}

In the following examples, we will use the trigger classes 10 and 14 and sub-groups of triggers found therein for illustration of the post classification analysis. 

\subsubsection{Auxiliary and Environmental channel connection}

Once the clusters of triggers are determined and the class members are assigned, the next question we ask is what are the possible couplings of these trigger clusters to the different sub-systems of the detector, i.e. what are the possible sources of these triggers? We tackle this problem by using the Omega scans~\cite{kleinewelle}. This information helps relate the cluster members to triggers seen in the auxiliary and environmental channels.

Omega scans are a set of time frequency plots that are based on logarithmic tiling of the
time–frequency plane that detect burst like signals in auxiliary and environmental channel data from the GW detectors. The first part is an application of
the dyadic wavelet transform and the the second is a somewhat modified windowed
Fourier transform that tiles the time–frequency plane for a specific Q-value. 

\begin{figure}
\includegraphics[scale=0.4]{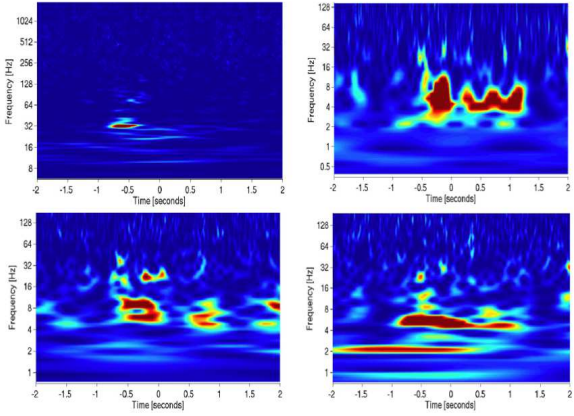}
\caption{
\label{omegascanfig}
The figure shows how a trigger in the GW channel and corresponding triggers in auxiliary and enviromental channels are seen in an omega scan. The top left panel shows a trigger in a GW channel between 16 and 32 Hz. Simultaneously, triggers are also seen in the end test mass and intermediate test masses in the X and Y arms of the interferometer (as seen in the other three panels).  The trigges seen in the auxiliary channles range in frequency between 8 and 32 Hz. The omega scans can be done on all available auxiliary and environmental channels that have been taking data at the time when the GW trigger happened. 
}
\end{figure}

Omega scans corresponding to triggers in a given class are generated and the corresponding auxiliary and environmental channels that showed triggers are noted. Figure~\ref{omegascanfig} shows an example of how a trigger in the GW channel and corresponding triggers in auxiliary and enviromental channels are seen in an omega scan. 

A cumulative list of auxiliary and environmental channels for each class of triggers is stored.

\begin{figure}
\includegraphics[scale=0.4]{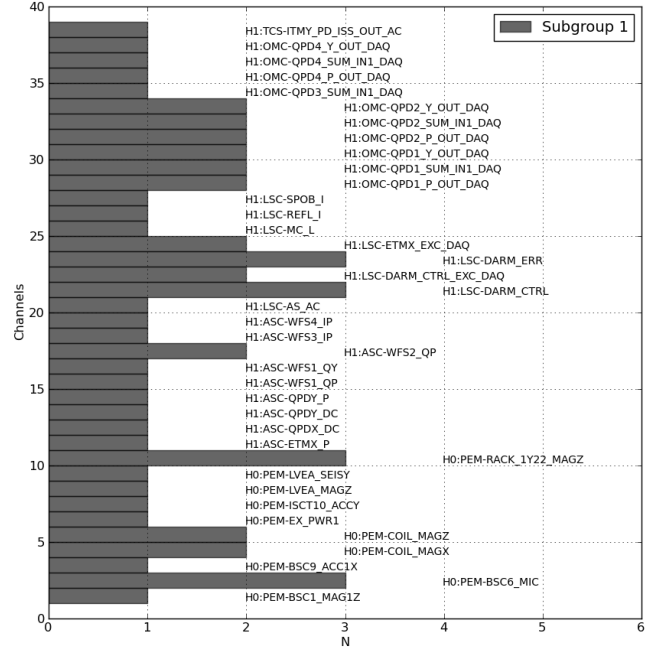}
\caption{
\label{class10group1}
The figure shows which auxiliary and environmental channels were seen to have triggered corresponding to the GW triggers seen in a certain class (\#10). As mentioned earlier, this class was seen to contain two statistically significant subclasses. This figure shows the auxiliary and environmental channels that triggered simultaneously for triggers in subclass 1.  }
\end{figure}

\begin{figure}
\includegraphics[height=23cm]{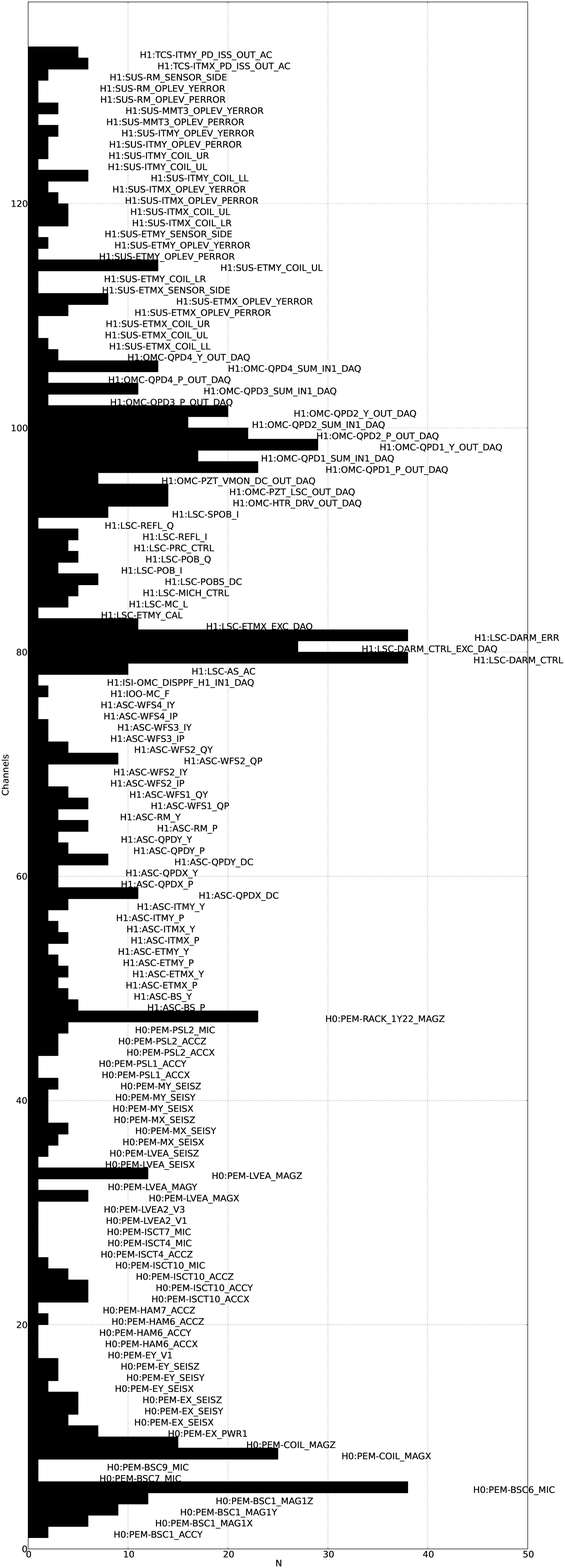}

\caption{
\label{class10group2}
The figure shows which auxiliary and environmental channels were seen to have triggered corresponding to the GW triggers seen in a certain class (\#10). As mentioned earlier, this class was seen to contain two statistically significant subclasses. This figure shows the auxiliary and environmental channels that triggered simultaneously for triggers in subclass 2.  }
\end{figure}

\begin{figure}
\includegraphics[scale=0.4]{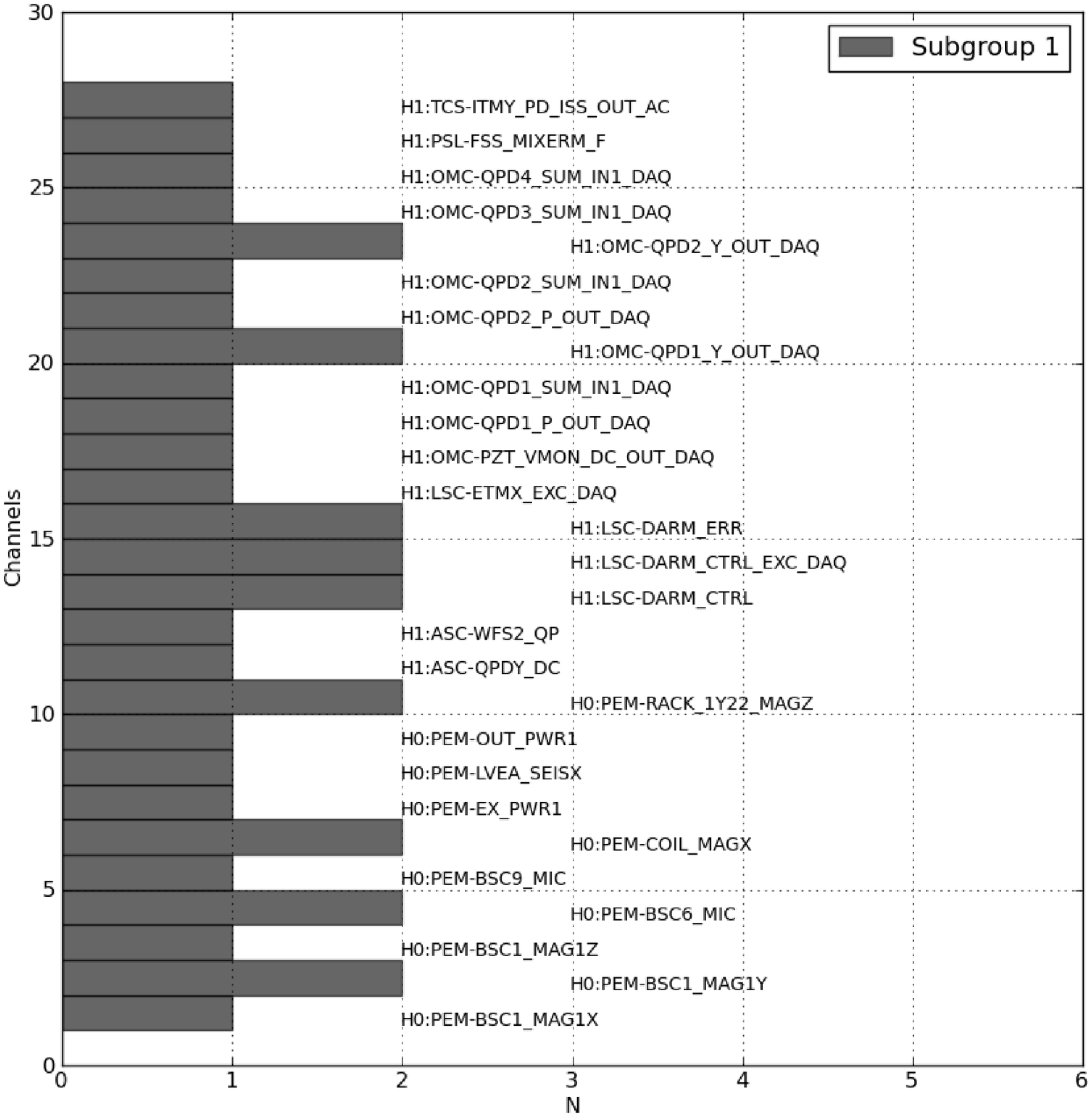}
\caption{
\label{class14group1}
The figure shows which auxiliary and environmental channels were seen to have triggered corresponding to the GW triggers seen in a certain class (\#14). As mentioned earlier, this class was seen to contain two statistically significant subclasses. This figure shows the auxiliary and environmental channels that triggered simultaneously for triggers in subclass 1.  }
\end{figure}

\begin{figure}
\includegraphics[height=23cm]{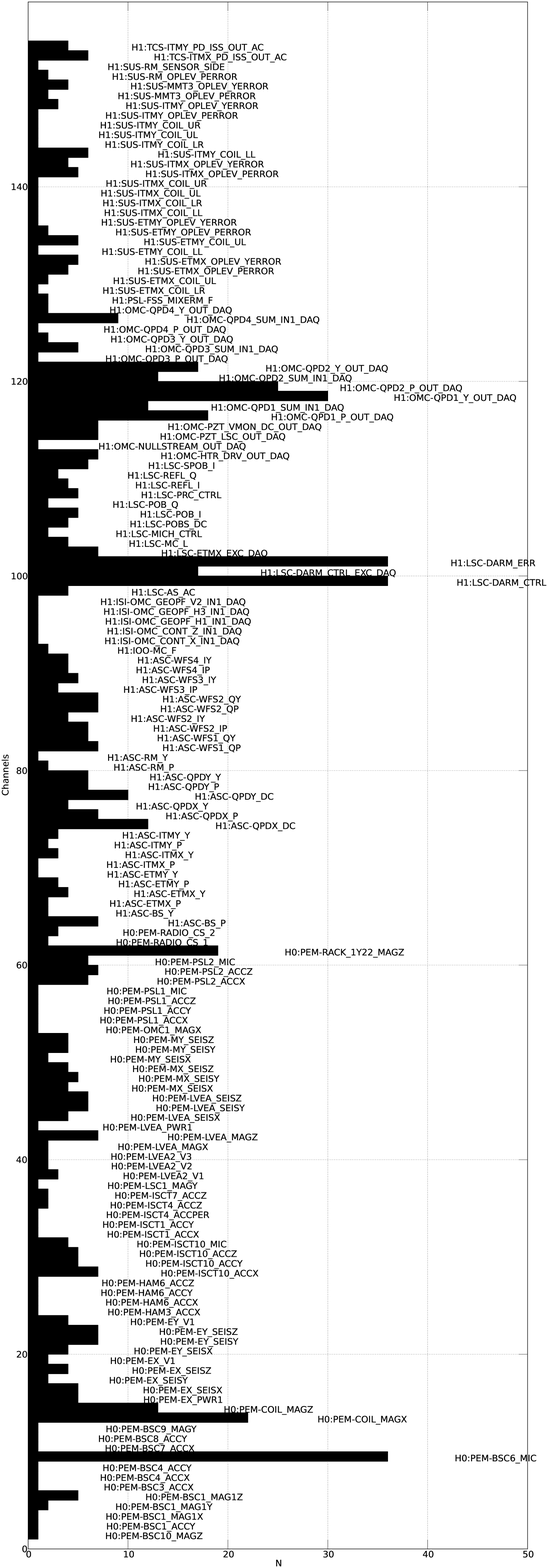}

\caption{
\label{class14group2}
The figure shows which auxiliary and environmental channels were seen to have triggered corresponding to the GW triggers seen in a certain class (\#14). As mentioned earlier, this class was seen to contain two statistically significant subclasses. This figure shows the auxiliary and environmental channels that triggered simultaneously for triggers in subclass 2.  }
\end{figure}

Figures~\ref{class10group1} and \ref{class10group2} show which auxiliary and environmental channels were seen to have triggered corresponding to the GW triggers seen in a certain class (\#10). Subclass 2 for this case was the larger of the two. Similarly, figures~\ref{class14group1} and \ref{class14group2} show which auxiliary and environmental channels were seen to have triggered corresponding to the GW triggers seen in another class (\#14).

\subsubsection{Comparison with existing data quality flags}
Data Quality (DQ) flags~\cite{DetChar, DQFlag} identify time periods in GW science data which are not suitable for astrophysical searches because of the varying statistical nature of the noise that is caused by instrumental malfunctioning in the detector and its surroundings. DQ flags are deemed effective if they can remove high snr glitches from the GW data streams. A large number of DQ flags exist within the LSC that are linked to various types of glitches and events observed in the data stream. We compare the auxiliary and environmental channel couplings observed in different classes in this study with those that already exist within the LSC repository.

\begin{figure}
\includegraphics[scale=0.42]{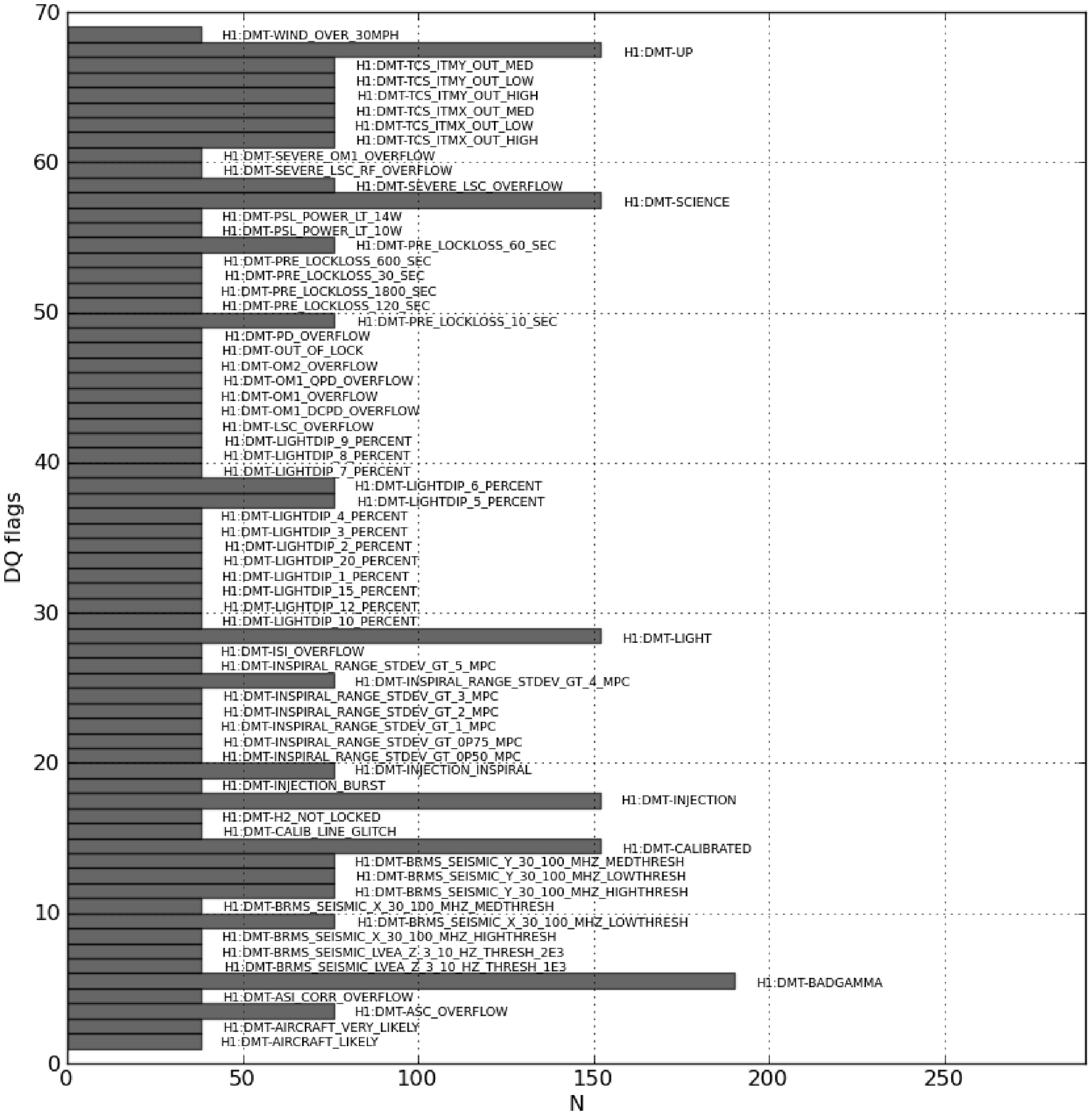}
\caption{
\label{class14dqflag}
The figure shows existing DQ flags corresponding to the GW triggers seen in class 14. The flags indicate that triggers in this class are relate to TCS glitches in the intermediate and end test masses in the X and Y arms of the interferometer, prestabilised laser power, pre-lock loss states, a few injections and glitches due to seismic reasons or flying aircrafts. }
\end{figure}

Figure ~\ref{class14dqflag} shows existing DQ flags corresponding to the GW triggers seen in class 14. These flags indicate that triggers in this class are related to TCS glitches in the intermediate and end test masses in the X and Y arms of the interferometer, prestabilised laser power, pre-lock loss states, a few injections and glitches due to seismic reasons or flying aircrafts. When compared to the couplings observed in the same class of triggers from the current study, we can see that a lot more information can be obtained from the coupled channels recorded in the current study. One distinction that can be readily made is that, the existing DQ flags being most often a result of human observation, are indicative of a qualitative cause rather than an exhaustive list of all auxiliary and environmental channels that might have caused the GW trigger. A combination of the existing data quality flags coupled with information from the LCSS+MHC trigger classes and their relation to the instrumental and environmental activities can furnish a more detailed and complete characterization of the triggers seen in GW channel.

\subsubsection{Characterization of the trigger classes}

Once all information as outlined above has been obtained, the trigger classes can be characterized in terms of (i) the waveform, (ii) range of physical properties, (iii) auxiliary and environmental channel couplings and (iv) DQ flags in use. Let us illustrate this using the class \# 10 and 14 as our example.  Table ~\ref{group_character} shows the mean values and range of snr, frequency and Q-values of the sub-groups found in classes 10 and 14 in our example. It is clear that the discriminating factor is the frequency. The reason that the main LCSS based class splits into sub-groups is because the high frequency triggers appear as outliers in the four dimensional hierarchical analysis. 

Following figures ~\ref{class10group1}, ~\ref{class10group2}, ~\ref{class14group1} and ~\ref{class14group2}, one can easily read off the auxiliary and environmental channel activities at the times of the occurrence of these triggers. 

Thus, the triggers in a given class and sub-group can be identified by the the shape of the trigger, the central frequency and the coupled channels and flags. The coupled channels can be ranked (in a statistical sense) by the percentage use ($\Lambda$) in a given day, as follows.

\begin{equation}
\Lambda = \frac{\Delta_{Channel}}{T_{Channel}} \times 100,
\end{equation}

where $\Delta_{Channel}$ referes to the fraction of triggers seen in a given auxiliary or environmental channel and $ T_{Channel}$ referes to the total number of triggers seen in all auxiliary and environmental channels. 
Table ~\ref{group_characterization} shows the $\Lambda$ values for channels coupled to the triggers in our example classes (\#10 and \#14). The percentage use ($\Lambda$) can thus serve as a pattern metric for each trigger group that comes with a characteristic waveform. In this particular example, it is quite evident that, apart from distinct shapes of the triggers belonging to the two groups, the top auxuliary and environmental channel percentage usages are different. While bth the classes do show very high percentage use for the channel OMC\_QPD (Output mode cleaner Quadrant Monitor Photodiode), class $\#14$ seems to have triggers caused (in a statistical sense) by the PEM (Physical environment monitor) MX and MY (Mid-station X and Y arms) Seismic (SEIS) activities that are not seen in the class $\# 10$ triggers. Seismic activities are recorded in the DQ flags ~\cite{DQFlag} that are being reported by other monitors. Another difference between the two classes is the presence of TCS (Temperature Control System) triggers in class $\# 14$, indicating that some of these triggers might have their origin in the TCS in the intermediate test masses in the X and Y arms (ITMX and ITMY) of the interferometer. These type of triggers are not found in class $\# 10$.

\begin{table}
\caption{\label{group_character}
This table shows the characteristics of trigger classes in terms of range of physical properties.
}
\begin{ruledtabular}
\begin{tabular}{c|c|c|c|c|}
Trigger class & N     &  snr &   frequency (Hz)  & Q-value \\ \hline \hline

Class 10 \\ Subgroup 1  & 32  &  26.4 $\pm$ 16.0  &  166.1 $\pm$ 128.9     &  29.7 $\pm$ 12.3\\ \hline
Class 10 \\ Subgroup 2  & 9    &  29.5 $\pm$ 15.9  &  1174.0 $\pm$ 435.9   &  21.6 $\pm$ 12.6\\ \hline 
Class 14 \\ Subgroup 1  & 2    &  33.9 $\pm$ 15.9  &  1777.0 $\pm$ 154.8   & 16.2 $\pm$ 3.0  \\ \hline
Class 14 \\ Subgroup 2  &  41 &  28.2 $\pm$ 17.2  &  331.2 $\pm$ 328.3     &  24.5 $\pm$ 8.8   \\ \hline
\end{tabular}
\end{ruledtabular}
\end{table}

\begin{table}
\caption{\label{group_characterization}
This table shows the $\Lambda$ values for channels coupled to the triggers in classes \#10 and \#14. A full description of the channels can be found in ~\cite{channels}. 
}
\begin{ruledtabular}
\begin{tabular}{c|c|c|c|}

Class \#14 couplings & $\Lambda$ \% & Class \# 10 couplings & $\Lambda$  \% \\ \hline 
OMC-QPD & 15.7  &  OMC-QPD & 24.4    \\ \hline
SUS & 7.3 &  PEM-COIL-MAG    &  6.2    \\ \hline
ASC-WFS & 6.95 & SUS-ETMX/Y & 5.4             \\ \hline
PEM-BSC & 6.82 &      OMC-PZT & 5.4         \\ \hline
ASC-QPDX & 5.9 &  ASC-WFS & 5.1       \\ \hline
PEM-EX/Y & 5.24 &  ASC-QPDX/Y & 5.0        \\ \hline
PEM-COIL & 4.6 &  PEM EX/Y & 4.6     \\ \hline
PEM-LVEA & 4.33 & PEM-BSC & 4.5         \\ \hline
PEM-ISCT & 3.8 &   SUS-ITMX/Y & 4.2       \\ \hline
PEM-MX/Y-SEIS & 3.01 &  PEM-LVEA & 3.7         \\ \hline
PEM-PSL & 3.01 &   PEM-ISCT & 3.2       \\ \hline
TCS-ITMX/Y & 1.31&  PEM-PSL & 1.9   \\ \hline
\end{tabular}
\end{ruledtabular}
\end{table}


\section{Conclusion and Future direction}
\label{future}

The study explores methods of time domain GW trigger classification using the shape parameters of trigger waveforms. The study extends to triggers noted in the GW channels as well as to all auxiliary and environmental channels. The classification into distinct groups is one of the most powerful data mining tools for analysis of large data sets, as is the case with LIGO science data.

Two algorithms have been tested here - (i) the LCSS and (ii) LCSS+FTSE, with the intent to test the relative computational speed and accuracy of classification. The different groups of triggers are indicators of certain common properties - in this case, similar types of waveforms - and thus can already reduce the dimensionality of the trigger identification problem by a large factor. This algoritm is then followed by MHC analysis. The integration of these two methods in a single analysis pipeline yields statistically significant classes of triggers with different waveform signatures and physical properties. These characteristics in turn are related to the processes that generate them and thus, classification of waveforms help shed light on very important aspects associated with tracking down trigger sources in the interferometer and its environment.

The current study was performed on simulated triggers in absence of noise and also in presence of various levels of noise to set benchmarks. The two algorithms differed in computational speed but no appreciable difference in performance in classifying the triggers accurately was noticed. The LCSS  and LCSS+FTSE  showed comparable computational speed for small samples (sample size $<$ 150). The combined FTSE +LCSS became rapidly more expensive with increasing sample size. The computation of LCSS from the intersection list of FTSE shown in figure \ref{Dyn_FTSELCSS} is carried out in three nested loops~\cite{MorsePatel}. The first loop is used to go to individual cells of the intersection list. The second is used to check individual values of a cell and the third one ensures the order of the subsequence which contribute to LCSS. The second and third loops could be avoided if their purposes could be taken care of while building the intersection list. The space complexity of LCSS+FTSE is very high compared to LCSS alone and most of the grid cells of the former are usually unoccupied. The amount of space could be reduced by increasing the threshold value~\cite{MorsePatel} (reducing the fineness of the grid) which, unfortunately would compromise the accuracy of LCSS. In the future applications of the combined FTSE and LCSS, the algorithms needs to be efficiently parallelized for treating large sample sizes. In case of triggers without noise, as is expected, both pipelines yielded an accurate classification structure, with each type of trigger being classified into the right cluster.

In case of triggers embedded in noise, the classification structure started to change from the true number of classes present in the data from snr $< $16. This happens because, with increasing noise, many of the characteristic trigger waveform features get masked by the mixed noise. This is illustrated in figure \ref{sim_noise_hist2} and explained in detail in section~\ref{results}. This is the reason that a careful trigger extraction pipeline has been employed in classification of the real LIGO S6 triggers. LIGO data is noise dominated, and sources of noise from auxiliary and environmental channels are many. A carefully constructed conditioning algorithm as described in section \ref{L4} ensures that the triggers are extracted with as much accuracy as possible, minimizing the noise content. This enhances the classification accuracy.

An important observation here is that classes produced based on similar waveforms can be heterogeneous in terms of its physical properties e.g. amplitude, Q-value, central frequency and snr of the triggers. Application of the MHC to look for further sub-classes in each of the waveform based classes is a very useful and productive step. This is the first time such an analysis is performed in the context of GW. As has been found in the study,the waveform based classes of triggers can be related to a group of auxiliary and environmental channels that are seen to have same type of triggers and the sub-classes can then further split the individual classes to  indicate which channels are most likely to be responsible for the production of these triggers. As stated in section \ref{results}, since each of the subgroups contain triggers with very characteristic properties and can be related to a specific set of channels, the method proves useful in classification of triggers seen in gravitational wave data and in helping with tracking down the sources or origins of the triggers. This can be a very effective means of characterizing the triggers and cataloging their properties and sources. 

The final outcome of the analysis shows existence of 19 statistically significant classes of triggers (in the data segments analyzed) with distinct waveforms and many more sub-classes with characteristic physical properties coming from GW channel, that can be coupled to different sets of auxiliary and environmental systems. Application of LCSS (which yields classes based on waveforms) alone would give 19 distinct classes, while application of only the MHC analysis would have given at most four statistically significant classes.

Over a sufficient length of time, enough knowledge base of trigger information can be compiled such that triggers can be identified without delay in a response to near real-time (low latency) needs of advanced LIGO's~\cite{ALIGO,ALIGO2, ALIGO3, Harry} detector characterization.


\begin{acknowledgements}
This work was supported by National Science Foundation grant PHY 0855371, 2009. LIGO Laboratory is acknowledged for usage of S6 data.
\end{acknowledgements}


\end{document}